\documentclass[11pt]{article}
\usepackage{amsmath}
\usepackage{amsfonts}
\usepackage[margin=.8in]{geometry}
\usepackage{graphicx}
\usepackage{subcaption}
\usepackage{float}
\usepackage{bm}
\usepackage{setspace}
\usepackage{natbib}
\PassOptionsToPackage{round}{natbib}

\begin{document}
\bibliographystyle{plainnat}
\doublespacing

\title{Methods for Bayesian Variable Selection with Binary Response Data using the EM Algorithm}
\author{Patrick McDermott, John Snyder, Rebecca Willison}
\maketitle

\begin{abstract}
High-dimensional Bayesian variable selection problems are often solved using computationally expensive Markov Chain Montle Carlo (MCMC) techniques. Recently, a Bayesian variable selection technique was developed for continuous data using the EM algorithm called EMVS. We extend the EMVS method to binary data by proposing both a logistic and probit extension. To preserve the computational speed of EMVS we also implemented the Stochastic Dual Coordinate Descent (SDCA) algorithm. Further, we conduct two extensive simulation studies to show the computational speed of both methods. These simulation studies reveal the power of both methods to quickly identify the correct sparse model. When these EMVS methods are compared to Stochastic Search Variable Selection (SSVS), the EMVS methods surpass SSVS both in terms of computational speed and correctly identifying significant variables. Finally, we illustrate the effectiveness of both methods on two well-known gene expression datasets. Our results mirror the results of previous examinations of these datasets with far less computational cost.
\end{abstract}

\section{Introduction}
The concurrent growth of Bayesian methodology and high dimensional datasets has created a need for high-dimensional Bayesian variable selection techniques. Since \cite{George1993} first introduced Stochastic Search Variable Selection (SSVS), many extensions have been developed for high-dimensional data. Many high-dimensional datasets, especially gene expression data, involve binary responses. Extensions to SSVS have also been developed throughout the literature for high-dimensional data with binary outcomes in \cite{Lee}, \cite{AI}, and \cite{Baragatti}. These Bayesian methods rely on MCMC methods that become computationally costly as the number of covariates grow. \cite{rockova} proposed a deterministic alternative using the EM algorithm, which is much faster than most SSVS methods. They present an EM variable selection (EMVS) scheme for continuous data.  The main objective of this paper will be to derive EMVS methods for binary response data with both logistic and probit models being considered.

The prior structure for our methods are based on the one used in \cite{GM97} to develop the SSVS method. We use the continuous conjugate versions of the ``spike-and-slab'' normal mixture formulation which induces a ``selective shrinkage'' property to aid in variable selection. The probit model is developed with the commonly used data augmentation technique of \cite{AlbertChib}. The recently developed Stochastic Dual Coordinate Ascent (SDCA) method of \cite{Shalev-A} can greatly improve computational times for regularized regression problems such as we have here. A SDCA method is utilized to both speed up the probit model and preserve the computational speed of EMVS for the logistic case.  Both of these methods are compared to a probit extensions of the SSVS model that uses a Gibbs sampler along with Metropolis steps from \cite{GM97}.  Additionally, the methods developed are available in an R package which can be accessed on the CRAN repository.

The rest of this article is organized as follows: Section 2 characterizes the hierarchical prior for EMVS, as well as the details for implementing the logistic and probit models. Section 3 provides two comprehensive simulation studies that compare both models with each other and to SSVS. In Section 4, we demonstrate both methods on two popular gene expression datasets. Finally, we end with a discussion of our extension of the EMVS technique to binary data in Section 5.
\section{Methods for Binary Data}
Here we examine both a logistic and probit extension of the original EMVS algorithm for binary data. A key feature of this extension concerns the two methods having nearly the same E-step, but several different derivations in the M-step. For both methods, suppose the response $\boldsymbol{y}$ is a  vector of length $n$ for $i,\ldots,n$; while $\boldsymbol{X}$ represents an $n\times p$ matrix of standardized predictor variables.
\subsection{Logistic and Probit likelihoods}
For the logistic model, we assume the data have outcomes $y_i \in \{\pm 1\} $ and we define $x_i \leftarrow -y_i x_i$. Further, we assume the following data likelihood:

\begin{equation}
\label{logisticlikelihood}
f(\boldsymbol{y}|\boldsymbol{\beta})=\prod_{i=1}^n \frac{1}{1+e^{\boldsymbol{x}_i^\prime\boldsymbol{\beta}}},
\end{equation}
where $\boldsymbol{\beta}$ is a vector of length p consisting of the regression coefficients for the model.\\
A popular alternative to the logistic model for analyzing binary data is the probit model.  In the probit model, our data is distributed $y_i\sim \text{Bernoulli}(p_i)$ where $p_i=\Phi\left(\boldsymbol{x}_i^\prime\boldsymbol{\beta}\right)$ and $\Phi(\cdot)$ is the cumulative distribution function of the standard normal distribution. Note, for the probit model we assume the data have outcomes $y_i \in \{0,1\} $. The data likelihood then becomes
\begin{equation}
\label{probitlikelihood}
f(\boldsymbol{y}\vert\boldsymbol{\beta}) = \prod_{i=1}^n \Phi(\boldsymbol{x}_i^\prime\boldsymbol{\beta})^{y_i}
(1-\Phi(\boldsymbol{x}_i^\prime\boldsymbol{\beta}))^{1-y_i}.
\end{equation}
\subsection{Spike-and-Slab Prior Structure}
As is the case with other Bayesian variable selection techniques, a length p vector $\boldsymbol{\gamma}$ of 1's and 0's is used for model selection, with the corresponding elements of $\boldsymbol{\gamma}$ signifying which columns of $\boldsymbol{X}$ are classified as being related to the response. Both models have the following hierarchical prior structure:
\begin{equation}
\pi(\boldsymbol{\beta},\sigma,\boldsymbol{\gamma})=\pi(\boldsymbol{\beta} \vert \sigma, \boldsymbol{\gamma})\pi(\sigma \vert \boldsymbol{\gamma})\pi(\boldsymbol{\gamma}) .
\end{equation}
with the prior on the regression coefficients following a ``spike-and-slab'' formulation as,
\begin{equation}
\pi(\boldsymbol{\beta} \vert \sigma, \boldsymbol{\gamma},\nu_0,\nu_1) \propto N_p\left(\boldsymbol{0},\boldsymbol{D_{\sigma,\gamma_i}}\right),
\end{equation}
where $\boldsymbol{D_{\sigma,\gamma_i}} = \sigma^2 \text{diag}(a_1,\ldots,a_p)$ and $a_i=(1-\gamma_i)\nu_0+\gamma_i \nu_1$.  This means that a variables exclusion or inclusion in the model sets its prior variance as $\nu_0$ or $\nu_1$ respectively.  This formulation was proposed by \citet{GM97}, and suggests setting the hyperparameters $\nu_0$ and $\nu_1$ to be small and large positive values, respectively.  Note that we assume that the intercept term has a uniform prior that has been marginalized out of the likelihood function, which is equivalent to centering the data at 0. The prior for $\sigma^2$ is conveniently set to be a noninformative Inverse Gamma prior, which is conjugate for the normal family:

\begin{equation}
\pi(\sigma^2 \vert \boldsymbol{\gamma},\lambda,\nu) \propto IG\left(\frac{\nu}{2},\frac{\nu \lambda}{2}\right),
\end{equation}
where $\nu=1$ and $\lambda=1$ to make the prior relatively noninformative.  Note that \citet{GM97} suggests that setting these hyperparameters as such suggest that $\lambda$ is the prior estimate of variance $\sigma^2$ and $\nu$ can be viewed as a prior sample size used to get that estimate. In formulating a prior for $\boldsymbol{\gamma}$, we will use the common Bernoulli-beta setup as:
\begin{align}
\begin{split}
\pi(\boldsymbol{\gamma} \vert \theta) \propto \ & \: \theta^{\sum_i{\gamma_i}} \left(1-\theta\right)^{p-\sum_i{\gamma_i}}, \\
\pi(\theta) \propto \ & \: \theta^{a-1}(1-\theta)^{b-1},
\end{split}
\end{align}
where $\theta$ can be viewed as the probability that any particular variable is in the model.

\subsection{EM Algorithm}
In order to estimate the unknown parameters $\boldsymbol{\beta}, \theta, \sigma,$ and $\boldsymbol{\gamma}$, the EM algorithm is used instead of the more popular MCMC method, because the algorithm offers substantial computational advantages when p is large.  In this implementation of the EM algorithm, the variable inclusion vector $\boldsymbol{\gamma}$ is used as the latent variable.  Since this variable cannot be observed, it will be replaced by its conditional expected value for the M-step, which will maximize the complete posterior with respect to $\Theta = (\boldsymbol{\beta}, \theta, \sigma)$. Recall that the EM algorithm maximizes the complete data log likelihood function with the latent variable replaced by its conditional expected value and the other parameters being replaced by their values from the previous iteration.  In other words, we will iteratively maximize the objective function
\begin{align}
\begin{split}
Q\left(\boldsymbol{\beta},\theta,\sigma \vert \boldsymbol{\beta}^{(k)},\theta^{(k)}, \sigma^{(k)}, \boldsymbol{y}\right) =
E_{\boldsymbol{\gamma} \vert .}   
\left[\log{\pi \left(\boldsymbol{\beta}, \theta, \sigma, \boldsymbol{\gamma} \vert \boldsymbol{y}\right)} 
\vert \boldsymbol{\beta}^{(k)},\theta^{(k)}, \sigma^{(k)}, \boldsymbol{y} \right].
\end{split}
\end{align}
The derivation of $\log{(\pi \left(\boldsymbol{\beta}, \theta, \sigma, \boldsymbol{\gamma} \vert \boldsymbol{y}\right)})$ is different for the logistic and probit model. The objective function $Q$ can be partitioned into two parts, $Q_1$ and $Q_2$ due to the conjugate nature of the hierarchical prior structure, so that:
\begin{align}
\begin{split}
Q\left(\boldsymbol{\beta},\theta,\sigma \vert \boldsymbol{\beta}^{(k)},\theta^{(k)}, \sigma^{(k)}, \boldsymbol{y}\right) =
C + Q_1\left(\boldsymbol{\beta},\sigma \vert \boldsymbol{\beta}^{(k)},\theta^{(k)}, \sigma^{(k)}\right)+
Q_2\left(\theta \vert \boldsymbol{\beta}^{(k)},\theta^{(k)}, \sigma^{(k)}\right)
\end{split}
\end{align}
 where C represents terms which are constant with respect to $\Theta$. Since the logistic and probit models have different likelihoods, the form of  $Q_1\left(\boldsymbol{\beta},\sigma \vert \boldsymbol{\beta}^{(k)},\theta^{(k)}, \sigma^{(k)}\right)$ is unique to each of the two models. One can show that for both models
\begin{align}
\label{q2}
Q_2\left(\theta \vert \boldsymbol{\beta}^{(k)},\theta^{(k)}, \sigma^{(k)}\right) =
\sum\limits_{i=1}^{p}{\log{\left(\frac{\theta}{1-\theta}\right)}}E_{\boldsymbol{\gamma} \vert .} \gamma_i +(a-1)\log{\theta}+(p+b-1)\log{(1-\theta)}.
\end{align}
This separability of $Q$ allows one to maximize the functions separately in the M-step.

\subsection{E-Step}
To compute the E-step, we must take the expectation of the aforementioned $Q$ function with respect to the latent variable $\boldsymbol{\gamma}$.  There are two parts of the formulations that need further examination, namely, $E_{\boldsymbol{\gamma} \vert .}\left[\frac{1}{\nu_0(1-\gamma_i)+\nu_1\gamma_i}\right]$ and $E_{\boldsymbol{\gamma} \vert .} (\gamma_i)$.  First, as \citet{rockova} discuss, the expectation of the latent variable depends on $\boldsymbol{y}$ only through $\left(\boldsymbol{\beta}^{(k)},\theta^{(k)}, \sigma^{(k)}\right)$ because of the hierarchical structure of $\boldsymbol{\gamma}$, therefore we have
\begin{align}
\begin{split}
E_{\boldsymbol{\gamma \vert .}}(\gamma_i) = P\left(\gamma_i=1 \vert \boldsymbol{\beta}^{(k)},\theta^{(k)}, \sigma^{(k)}\right)= p_i^* .
\end{split}
\end{align}
By an application of Bayes formula, we have
\begin{align}
\begin{split}
p_i^* = \frac{\pi(\beta_i^{(k)} \vert \sigma^{(k)},\gamma_i=1)P(\gamma_i=1 \vert \theta^{(k)})}
             {\pi(\beta_i^{(k)} \vert \sigma^{(k)},\gamma_i=1)P(\gamma_i=1 \vert \theta^{(k)}) +
              \pi(\beta_i^{(k)} \vert \sigma^{(k)},\gamma_i=0)P(\gamma_i=0 \vert \theta^{(k)})}.
\end{split}
\end{align}
Second, from the posterior distribution $\pi(\boldsymbol{\gamma} \vert .)$, we find the other conditional expectation by taking a weighted sum of the precision parameters
\begin{align}
\begin{split}
E_{\boldsymbol{\gamma} \vert .}\left[\frac{1}{\nu_0(1-\gamma_i)+\nu_1\gamma_i}\right] =
\frac{E_{\boldsymbol{\gamma} \vert .}(1-\gamma_i)}{\nu_0}+\frac{E_{\boldsymbol{\gamma \vert .}}\gamma_i}{\nu_1}=
\frac{1-p_i^*}{\nu_0}+\frac{p_i^*}{\nu_1}
\equiv d_i^* .
\end{split}
\end{align}
\subsection{Logistic Model}
If we use the Gaussian mixture prior on $\boldsymbol{\beta}$ and the hierarchical prior structure described above; EMVS for logistic regression begins with the E-step in Section 2.4. Next we derive the M-step for logistic regression with EMVS. Substituting in the log of (\ref{logisticlikelihood}), $Q_1$ yields the form
\begin{align}
Q_1=
 -\sum\limits_{i=1}^{n}\text{log}(1+e^{-x_i^\prime\boldsymbol{\beta}}) -\frac{p+\nu+2}{2}\log{\sigma^2} - \frac{\nu \lambda}{2\sigma^2} - & \frac{\sum{\beta_i^2}}{2\sigma^2}E_{\boldsymbol{\gamma} \vert .}\left[\frac{1}{\nu_0(1-\gamma_i)+\nu_1\gamma_i}\right].
\label{logQ1}
\end{align}
Differentiating $Q_1$ with respect to $\boldsymbol{\beta}$ reveals the following form for $\boldsymbol{\beta}^{(k+1)}$:
\begin{equation}
\label{logisticbeta}
\boldsymbol{\beta}^{(k+1)}=\text{arg} \min_{\boldsymbol{\beta}} \sum\limits_{i=1}^{n} \text{log}(1+e^{-x_i^\prime\boldsymbol{\beta}})+\| \boldsymbol{D}^{\boldsymbol{*}1/2}\boldsymbol{\beta}\|^2 ,
\end{equation}
where $\boldsymbol{D^{*}}$ is a $p \times p$ diagonal matrix such that $\boldsymbol{D^{*}}=\text{diag} \{ d_1^*,\cdots,d_p^*\}$. This is a difficult minimization problem due to the non-linearity associated with the first term in (\ref{logisticbeta}).

The regularized logistic model could be solved iteratively using the common Newton-Raphson numerical technique; however, in high dimensions the Newton-Raphson technique is known to be computationally expensive. Instead, we use the suggestion of \citet{rockova} and use a Stochastic Dual Coordinate Ascent (SDCA) algorithm from \citet{Shalev-A} to solve (\ref{logisticbeta}).

 From (\ref{logQ1}), one can show that for logistic regression
\begin{equation}
\sigma^{(k+1)}=\sqrt{\frac{\| \boldsymbol{D^}{\boldsymbol{*}1/2}\boldsymbol{\beta}^{(k+1)}\|^2 + \nu \lambda}{p+ \nu+2}} .
\end{equation}
To finish the specification of the M-step, one can show that from (\ref{q2}) that $\theta^{(k+1)}$ has the form
\begin{equation}
\label{theta}
\theta^{(k+1)}= \frac{\sum\limits_{i=1}^{n} {p_i^*}+a-1}{a+b+p-2} .
\end{equation}
\subsubsection{Stochastic Dual Coordinate Ascent Algorithm}
The advantage of the using a SDCA method to solve (\ref{logisticbeta}) over the Newton-Raphson technique lies in the computational speed of SDCA. We start by describing the general SDCA algorithm presented in \citet{Shalev-A}. Suppose we want to solve the following general regularization problem
\begin{equation}
\label{genopt}
\min_{\boldsymbol{w}\in \mathbb{R}^n} \ P(\boldsymbol{w})=\left[ \frac{1}{n}\sum\limits_{i=1}^{n}\phi_i(\boldsymbol{x}_i^\prime \boldsymbol{w})+\frac{\lambda}{2}\|\boldsymbol{w}\|^2\right],
\end{equation}
where $\phi_i$ is a scalar convex function. One should note that $\lambda$ is a scalar in (\ref{genopt}), whereas for our EMVS model, $\lambda$ is presented as a vector. The goal is to find a $\boldsymbol{w^*}$ that minimizes  (\ref{genopt}). For regularized logistic regression we have, $\phi_i(a)=\text{log}(1+\text{exp}(-y_ia))$. A common technique for solving regularized optimization problems involves finding the dual of the original function. If we let $\phi_i^*$  denote the convex conjugate of $\phi_i$, one can show the dual for $P(\boldsymbol{w})$ is:
\begin{equation}
\max_{\boldsymbol{\alpha} \in \mathbb{R}^n} \ \text{D}(\boldsymbol{\alpha})=\left[ \frac{1}{n}\sum\limits_{i=1}^{n}-\phi_i^*(-\alpha_i)+\left \| \frac{1}{\lambda n} \sum\limits_{i=1}^{n}\boldsymbol{x}_i^\prime \boldsymbol{\alpha}_i  \right \|^2\right].
\label{dualopt}
\end{equation}
In the case of regularized logistic regression $-\phi_i^*(-b)=-by_i\text{log}(by_i)+(1-by_i)\text{log}(1-by_i)$. Note that $\alpha_i$ is the dual for the $i$\textsuperscript{th} observation. This type of optimization method is often referred to as Dual Coordinate Ascent (DCA). DCA requires only one single $\alpha_i$ to be optimized at each iteration. The \citet{Shalev-B} method is stochastic because one randomly picks which $\alpha_i$  to optimized at each iteration. Finally, to complete the problem, define $w(\boldsymbol{\alpha})=\frac{1}{\lambda n} \sum\limits_{i=1}^{n} \boldsymbol{x}_i^\prime \boldsymbol{\alpha}_i$ and use the fact that $w(\boldsymbol{\alpha^*})=w^*$ and $\text{P}(\boldsymbol{w^*})=\text{D}(\boldsymbol{\alpha^*})$  where $\boldsymbol{w^*}$ and $\boldsymbol{\alpha^*}$ are the solution to (\ref{genopt}) and (\ref{dualopt}), respectively. The goal of SDCA is to find values of $\boldsymbol{\alpha}$ and $\boldsymbol{w}$ to minimize the so-called duality gap defined as $\text{P}(w(\boldsymbol{\alpha}))-\text{D}(\boldsymbol{\alpha})$. In order to use SDCA for logistic regression as described in \citet{Shalev-A}, one still needs to use a few Newton-Raphson steps at each M-step. \citet{Shalev-B}  introduced a sightly more efficient SDCA algorithm called Proximal Stochastic Dual Coordinate Ascent (Prox-SDCA) that avoids any Newton-Raphson steps. In the appendix, we present the Prox-SDCA algorithm used to solve (\ref{logisticbeta}).

\subsection{Probit Model}
The likelihood for the probit model (\ref{probitlikelihood}), is largely intractable, and becomes very complicated when we find the joint posterior distribution. For computational ease, we use the data augmentation method similar to the one proposed by \citet{AlbertChib} to introduce a latent variable $Z_i\sim N(\textbf{x}_i'\boldsymbol{\beta},1)$ such that
\[ Y_i = \left\{
  \begin{array}{l l}
    1, & \quad \text{if $Z_i>0$},\\
    0, & \quad \text{if $Z_i\le 0$}.
  \end{array} \right.\]
If we  observe $\boldsymbol{y}$ but $\boldsymbol{Z}$ is unobserved, then the distribution of $Z_i$ conditioned on $\boldsymbol{y}$ is truncated normal, with mean $\boldsymbol{x}_i^\prime\boldsymbol{\beta}$ and unit variance.  If $y_i=1$, then the distribution of $Z_i$ is truncated on the left by 0 (denoted by $p_1(\cdot)$), otherwise $y_i=0$ and the distribution of $Z_i$ is truncated on the right by 0 (denoted by $p_2(\cdot)$).  Note that if we observe $Z_i$, then $Z_i\sim N(\boldsymbol{x}_i^\prime\boldsymbol{\beta},1)$. The idea is to impute $\boldsymbol{z}$ so that we can then deal with normally distributed, continuous data rather than binary data.  This technique vastly improves computational complexity and efficiency.

Because we do not ``observe'' our data, we need to add a step to impute it during the E-step of each iteration of the EM algorithm.  We estimate $\textbf{z}^{(k)}\vert\boldsymbol{x},\boldsymbol{y},\boldsymbol{\beta}$, by finding
\[z_i^{(k)}=E[z_i\vert \boldsymbol{x}, \textbf{y},\boldsymbol{\beta}^{(k)}] = \left\{
  \begin{array}{l l}
    \boldsymbol{x}_i'\boldsymbol{\beta}^{(k)}+ \frac{-\phi(-\boldsymbol{x}_i^\prime\boldsymbol{\beta}^{(k)})} {\boldsymbol{\Phi}(-\boldsymbol{x}_i^\prime\boldsymbol{\beta}^{(k)})}, & \quad \text{if $y_i=0$}\\
    \boldsymbol{x}_i^\prime\boldsymbol{\beta}^{(k)}+ \frac{\phi(-\boldsymbol{x}_i^\prime\boldsymbol{\beta}^{(k)})} {1-\boldsymbol{\Phi}(-\boldsymbol{x}_i^\prime\boldsymbol{\beta}^{(k)})}, & \quad \text{if $y_i=1$}.
  \end{array} \right.\]

This follows from the properties of the truncated normal distribution. Now that we have estimates for $\boldsymbol{z}^{(k)}\vert\boldsymbol{x},\boldsymbol{y},\boldsymbol{\beta}$, we can treat them as the data during each iteration of the EM algorithm so that the objective function becomes
\begin{align}
\begin{split}
Q\left(\boldsymbol{\beta},\theta,\vert \boldsymbol{\beta}^{(k)},\theta^{(k)}, \boldsymbol{z}^{(k)}\right) =
E_{\boldsymbol{\gamma} \vert .}   
\left[\log{\pi \left(\boldsymbol{\beta}, \theta,  \boldsymbol{\gamma} \vert \boldsymbol{z}^{(k)}\right)} 
\vert \boldsymbol{\beta}^{(k)},\theta^{(k)},  \boldsymbol{z}^{(k)} \right].
\end{split}
\end{align}

With the above objective function in hand, one can derive the probit E-M algorithm with relative ease.  There are two different methods for estimating $\boldsymbol{\beta}^{(k+1)}$ in the M-step.  The more traditional generalized ridge regression solution could be used, as well as a more computationally efficient solution such as the SDCA algorithm for continuous data from \citet{Shalev-A}.  Our specific E and M steps for the probit model are detailed in the appendix.

\section{Simulation}
\subsection{Method for Simulation of Data}
To generate a design matrix, we use the method of \citet{GuiLi2005}.  This method allows us to consider a synthetic data set where $p \gg n$, but a very small number of variables, $p_\gamma$, are actually related to the survival time.  The remaining $p-p_\gamma$ variables are not related to the survival time, but can be correlated with the related ones. \\

\hspace{4ex} The process, as described in \citet{GuiLi2005}, starts by generating an $n\times p$ matrix $A$ of $U(-1.5,1.5)$ random variables. We then take the first $p_\gamma$ columns in this matrix as the related variables that will be used to generate response data.  To generate the unrelated variables, we start by obtaining a normalized basis of $A$ using Gram-Schmidt orthonormalization, which we denote as
$\left\{\upsilon_1,\ldots,\upsilon_{p_\gamma},\varrho_1,\ldots,\varrho_{n-p_\gamma}\right\}$.

An application of Cauchy's inequality can be used to show that for any $p_\gamma \times (n-p_\gamma)$ matrix $\boldsymbol{T}$,
\begin{equation}
\label{corr}
\rho = corr\left(\boldsymbol{\upsilon}y,\left(\boldsymbol{\varrho}+\boldsymbol{\upsilon} \boldsymbol{T}\right)x\right) \le
\frac{\lambda}{\sqrt{1+\lambda^2}},\, \forall y\in \mathbb{R}^{p_\gamma}, \forall x\in \mathbb{R}^{n- p_\gamma},
\end{equation}
where $\lambda^2$ is the largest eigenvalue of $\boldsymbol{T}^\prime\boldsymbol{T}$. \newline
\indent We then find $\lambda^2 = \frac{\rho^2}{1-\rho^2}$ , where $\rho$ is the desired maximum correlation between the related and unrelated variables. Finally, $\boldsymbol{T}$ is then constructed as a diagonal matrix with its largest value as $\lambda$, and the remaining $p-p_\gamma$ variables are generated from the linear space $C=\left\{ \boldsymbol{\varrho}+\boldsymbol{\upsilon} T \right\}$.

\subsection{Simulation for binary data}
Using the method described in Section 3.1, a design matrix $\boldsymbol{X}$ is generated with 0.6 chosen as the maximum correlation between related and unrelated variables. Suppose we set $\boldsymbol{\beta}=(1,2,3,0,0,\ldots,0)'$ so only the first 3 variables are significant and $\epsilon\sim \text{N}_n(\boldsymbol{0}, \sigma^2_{\epsilon} \text{I}_n)$ with $\sigma^2_{\epsilon}=3$. Then, we can generate continuous responses through $\boldsymbol{y}=\boldsymbol{X}\boldsymbol{\beta} + \epsilon$.  We transform the aforementioned continuous vector $\boldsymbol{y}$ for the logistic and probit models. The continuous outcomes are transformed using the transformation $p_i=1/(1+\text{exp}(-y_i))$. The binary outcome is defined such that, P$(y_i=1)=p_i$ for both logistic and probit, while for logistic, P$(y_i=-1)=1-p_i$, and for probit, P$(y_i=0)=1- p_i$. For both methods we varied $\nu_0$, while fixing $\nu_1=1000$ in the logistic case and $\nu_1=100$ in the probit case. Further, we used the prior $\theta \sim \text{U}(0,1)$, which leads to a beta-binomial prior for $\boldsymbol{\gamma}$. Similar to \citet{rockova}, we set the starting values $\sigma^{(0)}=1$ and $\theta^{(0)}=.5$ .  The hyper-parameter $\lambda$ is set to .001 for the logistic model while $\nu$ is set to 1. Since the model is extremely sparse, we use the suggestion from \citet{rockova} of setting $a=1$ and $b=p$ . For sparse settings such as this, $b=p$ appears to produce more accurate results.

\begin{figure}[H]
\centering
\includegraphics[scale=0.5]{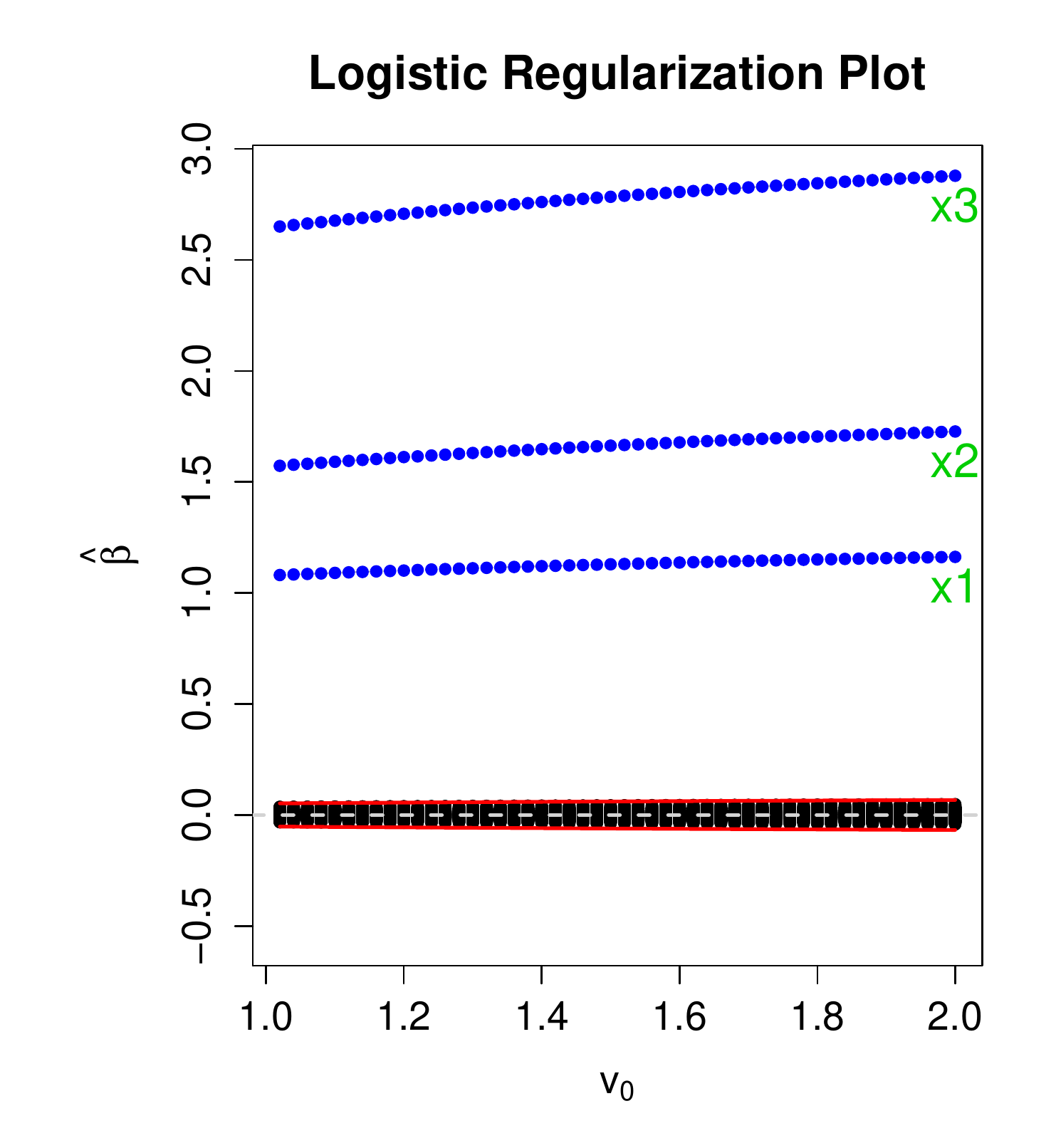}
\caption{Using EMVS for logistic Regression method, values of estimated regression parameters are plotted across different values of $\nu_0$. Only the first three parameters that were set as significant in the true model are found to be significant by the EMVS model. }
\label{logisticreg}
\end{figure}

\begin{figure}[htbp]
    \centering
    \begin{subfigure}[H]{0.4\textwidth}
        \includegraphics[width=2.75in]{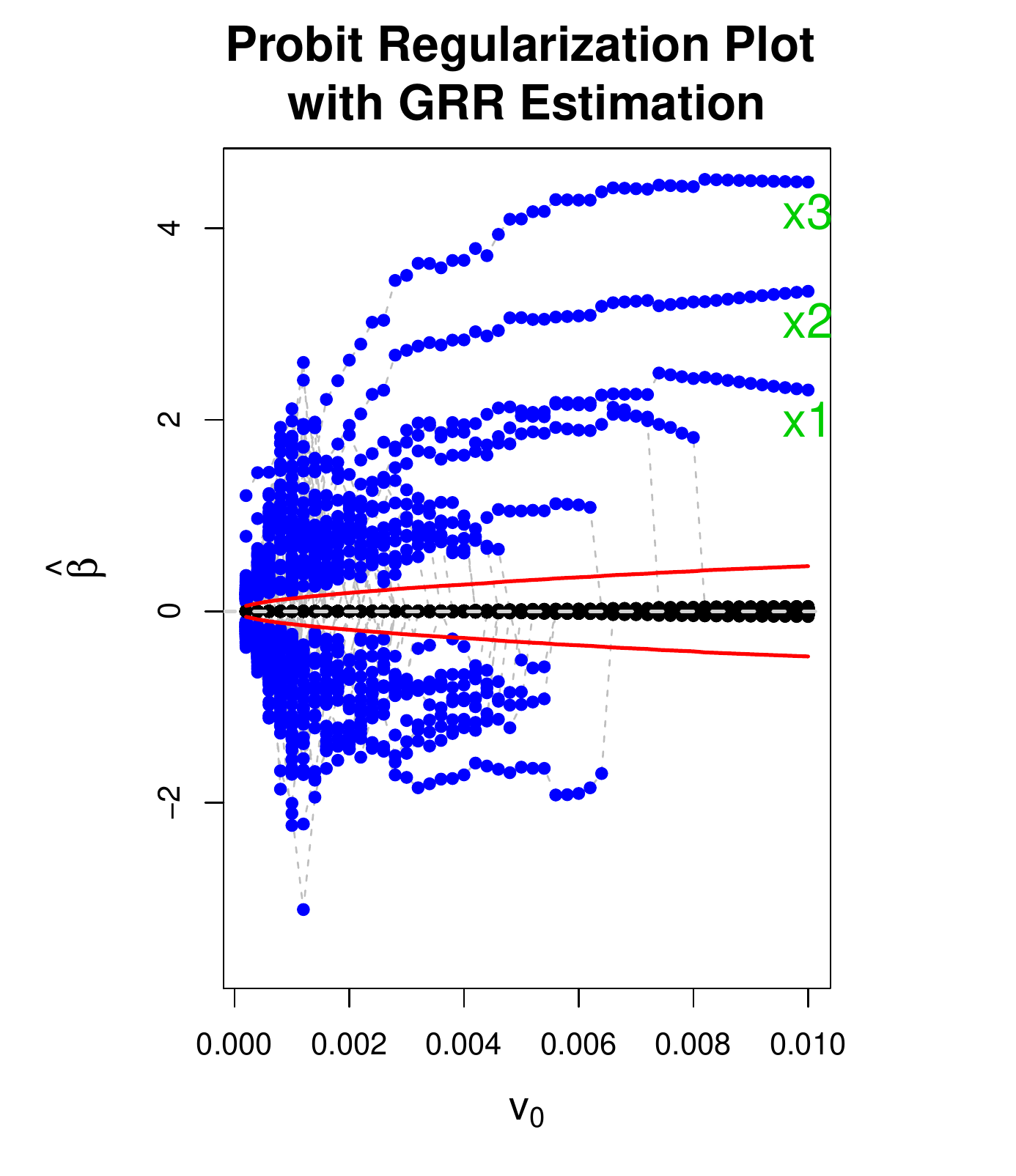}
        \caption{}
        \label{}
    \end{subfigure}
    ~ ~
    \begin{subfigure}[H]{0.4\textwidth}
        \includegraphics[width=2.75in]{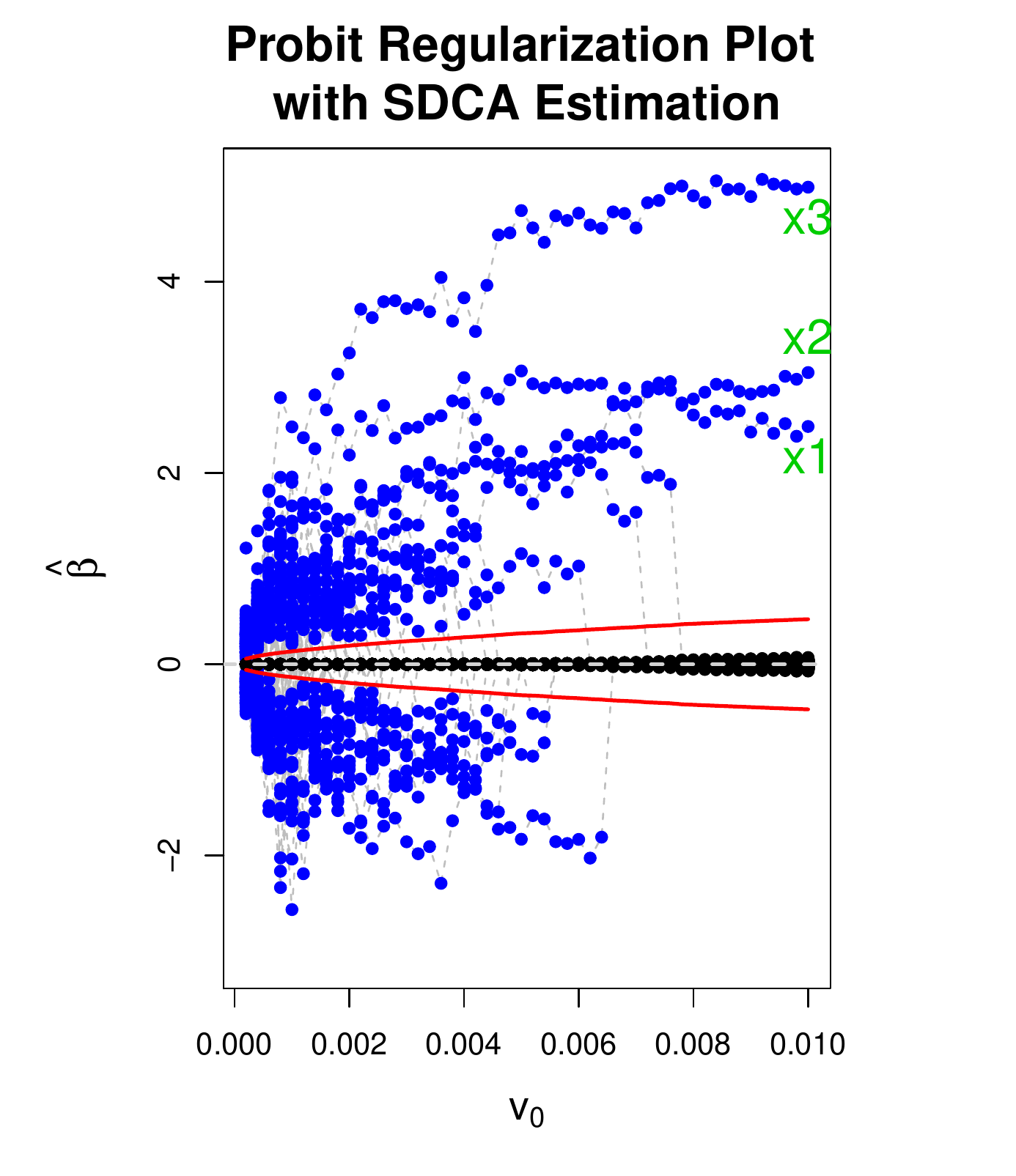}
        \caption{}
        \label{}
    \end{subfigure}
    \caption{(a) Uses GRR estimation to estimate regression parameters for the probit model. Estimates are relatively stable across different values of $\nu_0$. (b) uses a SDCA algorithm to estimate parameters for the probit model, leading to more volatile estimates across different values of $\nu_0$.}
    \label{probitreg}
\end{figure}

 We are able to set the number of iterations to $K=5,000$ due to the speed of the SDCA and Prox-SDCA algorithms. Setting $K$ any higher results in very slight gains in accuracy. In practice, values of $K$ will be application dependent. Overall, EMVS for logistic regression has fast computational times due to the use of Prox-SDCA. The whole algorithm typically converged in under 10 seconds on a quad core 3.4GHz i7 with 8GB of RAM running Windows 7. We found estimating $\boldsymbol{\beta}$ with the SDCA algorithm resulted in similar run times for the probit model. As shown in Figure \ref{logisticreg} and Figure \ref{probitreg}, we ran both models over a grid of 50 different $\nu_0$ values. The values of $\nu_0$ ran from 1.02 to 2 by increments of .02 for the logistic model; while for the probit model, values of $\nu_0$ stretched from .0002 to .01 with increments of .0002. Recall that $\gamma_i$ is the posterior probability of the $i^{\text{th}}$ variable being included in the model. Note that in Figure \ref{logisticreg} and Figure \ref{probitreg}, variables that have a $\gamma_i$ less than .5 are shown between the two red lines.  We found that in general, the logistic model preformed better for larger values of $\nu_0$ while the probit model preformed better for smaller values. Figure \ref{probitreg} portrays the $\boldsymbol{\beta}$ estimates for the probit model with both the GRR estimator and the SDCA algorithm. With $p=1000$ and the number of iterations for the SDCA algorithm equal to 6000, the GRR estimator and the SDCA algorithm produce similar results with comparable run times.  As one increases $p$, the SDCA algorithm becomes increasingly computationally efficient compared to the GRR estimator.  One option for getting starting values for the EMVS algorithm involves using a regularized logistic estimator.  In particular, we could use the Prox-SDCA Algorithm above to get robust starting values for logistic regression. This strategy was employed to produce Figure \ref{logisticreg} and Figure \ref{probitreg}.
\vspace{.50cm}
\vspace{-.75cm}
\subsection{Comparison with Stochastic Search Variable Selection}
To demonstrate the power of our EMVS methods, we compared our results to a SSVS model calibrated for binary data. For the sake of comparison, we used the same conjugate spike-and-slab prior for the SSVS model. We closely followed the Metropolis algorithm from \cite{GM97} that also used a conjugate prior structure. Conveniently, under the conjugate spike-and-slab prior the marginal posterior for $\boldsymbol{\gamma}$ does not depend on $\boldsymbol{\beta}$, hence for the continuous case one only needs to sample $\boldsymbol{\gamma}$. For binary data we once again use the data augmentation technique of \citet{AlbertChib} to work with continuous latent variables. In order to sample the latent variables, we must also sample $\boldsymbol{\beta}$. Thus, the SSVS algorithm becomes a Gibbs sampler with Metropolis steps to find $\boldsymbol{\gamma}$. Further, the data is simulated exactly the same as described in Section 3.2. The same beta-binomial prior is used as above with $\theta \sim U(0,1)$, along with setting $\nu_0=0$ and $\nu_1=1000$. We set the number of iterations equal to the time it took the EMVS model for logistic regression to go through 50 values of $\nu_0$. The SSVS model only got through 700 iterations of the entire Gibbs sampler and performed 700,000 iterations of the Metropolis algorithm in the time it took the EMVS method to complete. The SSVS had an acceptance rate of .006. Recall, we generated the data with $\boldsymbol{\beta}=(1,2,3,0,0,\ldots,0)'$. The SSVS technique selected a model that only included the predictor $\{\beta_3\}$ opposed to both the logistic and probit with EMVS which selected a model that included all 3 significant predictors, $\{\beta_1,\beta_2,\beta_3\}$. Thus, the developed EMVS techniques for binary data were both faster and better at correctly identifying the significant variables than the SSVS methods.
\subsection{Simulation Study}

While the simulation studies in Section 3.2 and 3.3 illustrate the power of our EMVS methods for binary data, the selected \textquotedblleft true\textquotedblright $\boldsymbol{\beta}$ values were rather arbitrary. To further study our developed methods, we created a second simulation which produced a variety of realistic datasets with less arbitrary true $\boldsymbol{\beta}$ values.  We used the following algorithm for the second simulation:
\begin{enumerate}
\setlength\itemsep{-.25em}

\item Fix $n=100, p=1000, p_{\gamma}=10, \rho=.6, \sigma^2_{\epsilon}=3,$ and $\beta_{max}$.

\item Generate an $n \times p$ design matrix $\mathbf{X}$ using the method outlined in 3.1, such that $p_{\gamma}$ variables will be related to the response and have maximum correlation $\rho$ with the remaining $p-p_{\gamma}$ variables.

\item Generate $p_{\gamma}$ true coefficients from $UNIF(-\beta_{max},\beta_{max})$, and set the remaining $p-p_{\gamma}$  coefficients to 0.

\item Generate continuous response data and convert this to binomial data.

\item Evaluate the probit and the logit methods with respect to True Positive Rate, True Negative Rate, Positive Predictive Value, and Negative Predictive value.
\item Repeat steps 2-5 500 times.
\end{enumerate}

This method was done for $\beta_{max}$ values of 0.5, 1, 2, and 3.  The results indicated that the logistic method excels in with larger values of $\nu_0$, while probit excels with smaller values, confirming our earlier observation and mirroring performance in real applications.  The plot for $\beta_{max}=2$ is shown below.  The plots for the other values of $\beta_{max}$ are in the Appendix and show similar patterns, only shifting vertically depending on the magnitude of $\beta_{max}$.

\vspace{-15mm}

\begin{figure}[H]
\centering
\includegraphics[scale=0.5]{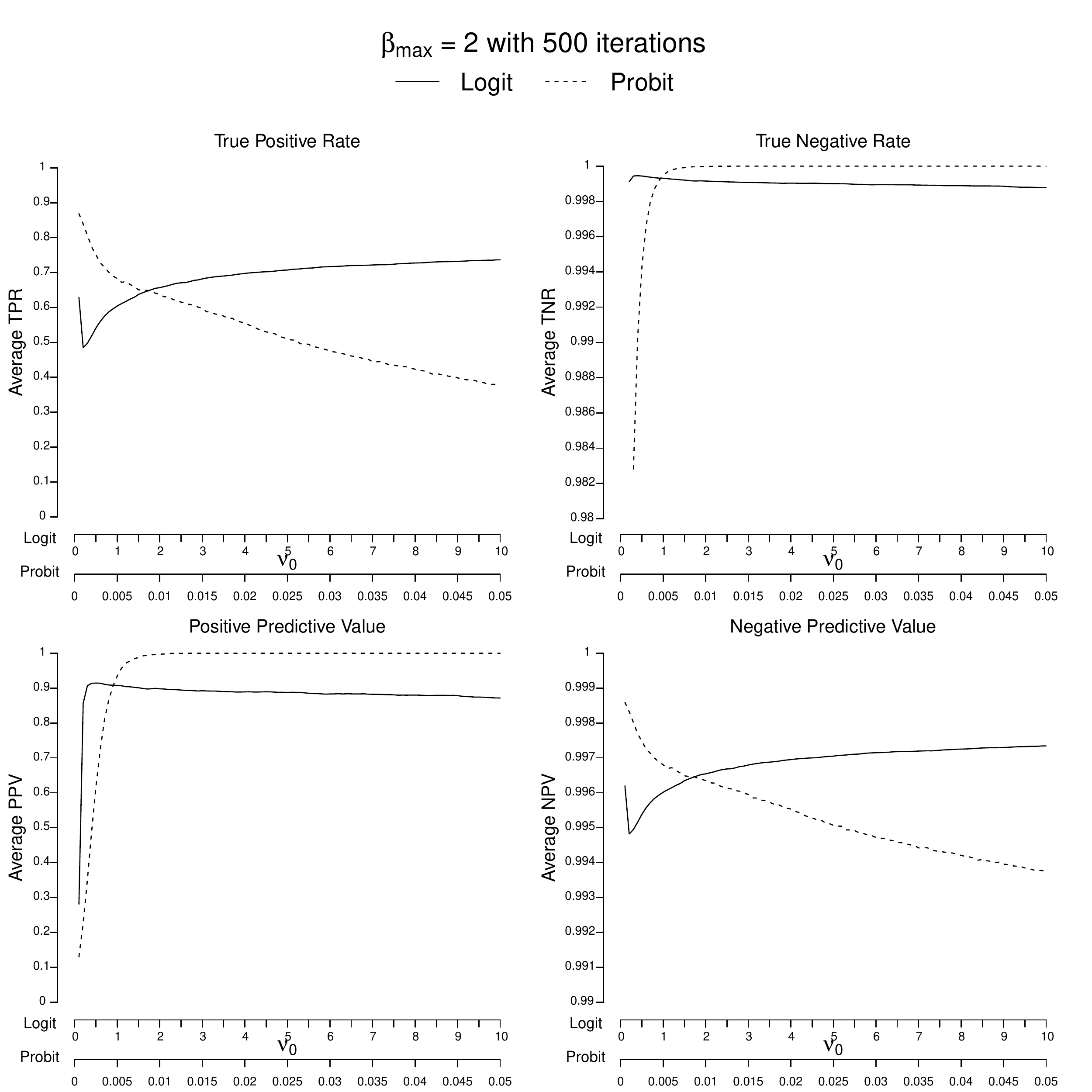}
\caption{TPR, TNR, PPV, and NPV for the logit and probit models over the tuning parameter $\nu_0$ for 500 simulated datasets using a maximum effect size of 2.}
\label{protbeta2}
\end{figure}

Other than as a demonstration of where each data model performs best with respect to the tuning parameter $\nu_0$, these plots indicate that the probit model, with its lower true positive rate and almost perfect positive predictive value, is a more conservative model for binary data.  The logit model, on the other hand, has a higher true positive rate but a lower positive predictive value, meaning we can be less certain a variable is actually related to the response should it be selected by the method.

\section{Applications}
\subsection{Logistic Application to Leukemia Data}
The well studied leukemia dataset from \citet{Golub} provides a good benchmark to evaluate the EMVS algorithm with logistic regression. We also analyzed the results from \citet{Lee}, which used the same leukemia dataset as \citet{Golub} with a Bayesian stochastic search algorithm. \citet{Lee} provides an ideal comparison since the usefulness of the EMVS algorithm lies in its computational speed compared to other Bayesian variable selection techniques.  For comparison purposes, we used the same 38 sample training data set that \citet{Lee}  and \citet{Golub} analyzed. Further, the design matrix was centered and scaled. The leukemia dataset contains 7,129 genes with the binary responses acute myeloid leukemia (AML) or acute lymphoblastic leukemia (ALL). We used the same hyper-parameters as described in Section 3.2, as well as the Prox-SDCA algorithm to get starting values. For the Prox-SDCA algorithm, setting the number of iterations equal to 10 times the number of samples, T=380, produced optimal results. With such a low number of iterations, the whole algorithm takes only 30 seconds to complete. Overall, this is a huge improvement in computational time compared to previous methods. Specifically, \cite{AI} reported that their SSVS analysis of this dataset took 282 minutes. Genes that had posterior probabilities greater that .5 were considered significant. In Table \ref{genetable}, 18 of 32 genes that were identified as significant were also identified by either \citet{Golub} or \citet{Lee}.
\small
\begin{table}[H]
\centering

\begin{tabular}{c c c c c}
\hline
Frequency ID & Gene ID & Frequency ID &  Gene ID  &  \\ [0.5ex] 
\hline
461 & D49950 &  1249 & L08246**  \\
1745 & \bf{M16038} & 1779 & M19507 \\
1829 & M22960 & 1834 & \bf{M23197} \\
1882 & \bf{M27891} & 2020 & M55150**  \\
2111 & M62762** & 2121 & M63138** \\
2181 & M68891 & 2242 & M80254** \\
2267 & M81933 & 2288 & \bf{M84526}  \\
2402 & M96326* & 3258 & U46751** \\
3320 & U50136**  & 3525 & U63289 \\
3847 & U82759** & 4052 & X04085**\\
4196 & X17042** & 4229 & X52056 \\
4377 & X62654 & 4499 & X70297 \\
4847 & X95735* & 5039 & Y12670** \\
5954 & Y00339 & 6041 & L09209* \\
6376 & M83652** & 6539 & X85116 \\
6677 & X58431 & 6919 & X16546\\
\hline

\end{tabular}
\label{genetable}
\caption{EMVS with logistic regression for $\nu_0=7$. The stared(*) genes were also identified as significant in \citet{Lee}, while the double stared(**) genes were identified as significant in \citet{Golub}. The bolded genes were recognized as significant in both papers. }\label{genetable}
\end{table}
\normalsize
To analyze the predictive power of the model, we randomly divided the full original leukemia data into a test and training dataset. The original dataset, before it was split, had 72 samples. Further, 48 observations were randomly selected for the training dataset while the remaining observations were left for the test dataset. The data was pre-processed as described in \citet{Dudoits}. After pre-processing, the dataset now contained 3,571 genes. All 3,571 genes  were used to produce predictions. Using a cut-off of .5, the model correctly predicted 23 out of the 24 observations from the test data set. \citet{Lee} found similar results with their stochastic variable selection model, although they only used the significant variables to produce predictions.

\subsection{Probit Application to Colon Cancer Data}
Another well-studied data set is the colon cancer gene expression data set collected by \citet{Alon}.  In the study, 40 tumor and 22 normal colon tissues were analyzed with Affymetrix oligonucleotide array to extract useful gene expression patterns from over 6,500 genes.  \citet{Alon} then selected a subset of 2,000 genes based on the confidence in expression levels.  This data has been analyzed extensively in the literature using many different statistical methods, including clustering \citep{Alon,BenDor,Li}, support vector machines \citep{BenDor,Furey}, boosting \citep{BenDor}, partial least squares \citep{Nguyen}, and evolutionary neural networks \citep{Kim}.  Nearly all of these studies used some sort of method of feature selection to reduce noise and improve prediction, so comparison between methods is somewhat limited. Usefulness of gene selection prior to analysis is heavily dependent on the method used for prediction. Additionally, five tissues that were suspected to be contaminated (N34, N36, T30, T33, and T36)\citep{Alon,Li} were removed from analysis.

The gene expression levels were log-transformed prior to analysis, and a repeated random sub-sampling validation scheme was used.  Thirty random samples were constructed, in which 20\% of the observations were held out, and then the model was fit using the remaining 80\% of each sample. Prediction accuracy was averaged over these 30 subsamples, and the best identified genes were those identified most frequently from the 30 models.  A gene was considered ``significant'' if the posterior probability was greater than 0.5, although posterior probabilities tended to be very close to 0 or very close to 1.  Table \ref{probittable} below shows the most commonly identified genes.
\small
 \begin{table}[ht]
\centering

\begin{tabular}{c c c c c}
\hline
Frequency ID & Gene ID & Frequency ID &  Gene ID  &  \\ [0.5ex] 
\hline
245 &M76378* & 897 &H43887* \\
249 &M63391* & 1042 & R36977* \\
267 &M76378* & 1058 &M80815* \\
365 & X14958*& 1325 &T47377*\\
377 & Z50753*& 1423 &J02854\\
493 & R87126*& 1582 &X63629*\\
561 &R46753& 1635 &M36634*\\
625 & X12671*& 1771 &J05032*\\
698 &T51261& 1772 &H08393*\\
780 & H40095*&1836& U14631 \\
802 & X70326*& 1870 & H55916*  \\
822 &T92451*& 1884 &R44301  \\
892 &U31525& 1894 &X07767\\
\hline
\end{tabular}
\label{probittable}
\caption{Genes with (*) were identified by other methods.}\label{probittable}
\end{table}
\normalsize
\begin{figure}[H]
\centering
\includegraphics[scale=0.5]{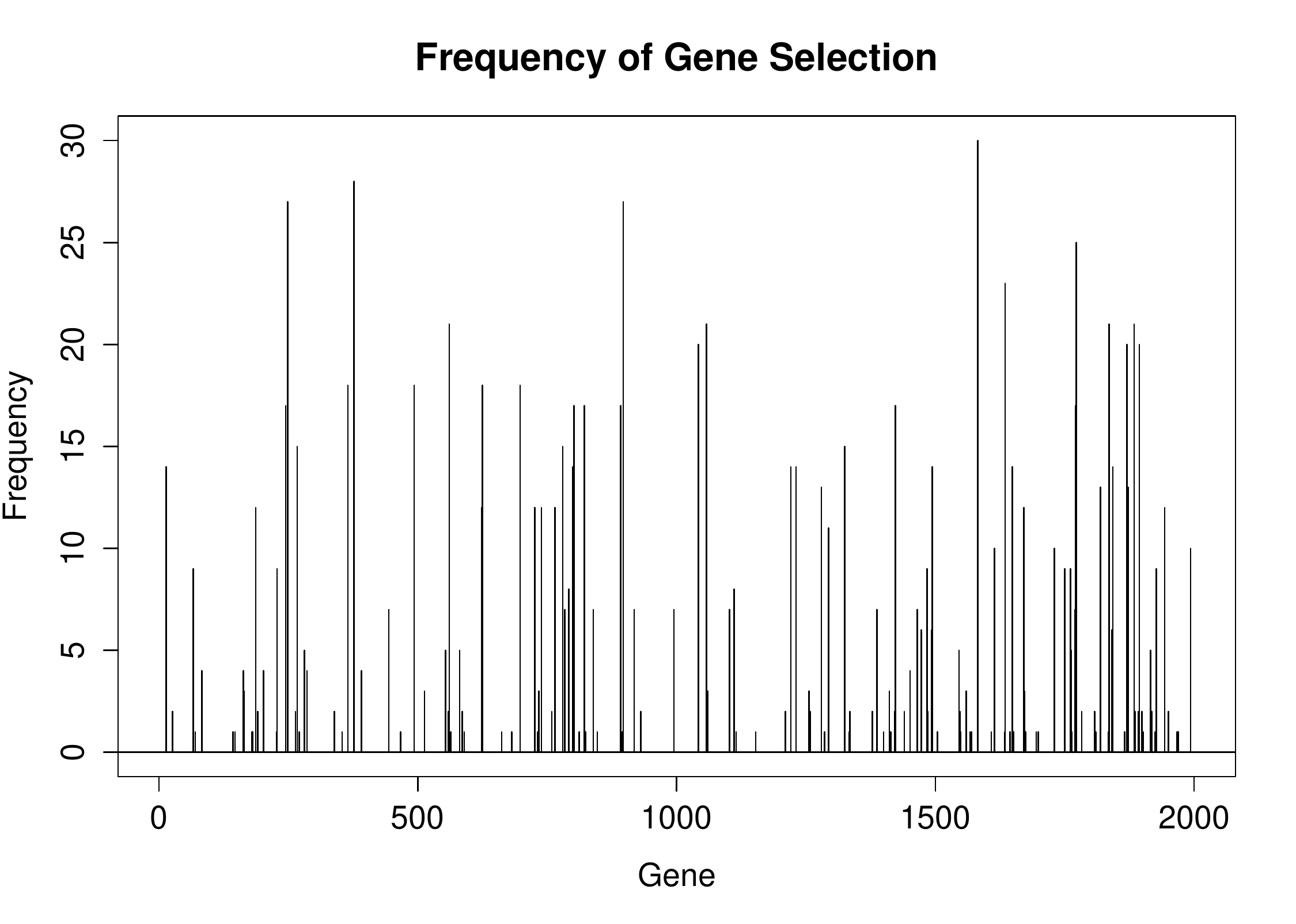}
\caption{This plot shows the frequency with which genes were selected when fitting the model to 30 subsamples of the data.}
\end{figure}

Many of the commonly identified genes have also been identified in other studies such as \citet{BenDor,Kim}.  Table \ref{probittable} lists the top 26 genes, but most of the 111 genes identified by EMVS had been identified by other papers as well.  A nice feature of data analysis using EMVS is that there is a clear distinction between variables which should be kept in the model as posterior probabilities are either very close to 1 or 0.  Most other feature selection methods for this data set have used various scoring procedures with a subjective cutoff used to select the ``top'' genes.  With EMVS, there is no ambiguity as to which variables should be used.

Additionally, for each of the 30 random subsamples used for cross-validation, prediction was very accurate.  The average prediction success rate was 93\%; the range of success rates using other methods ranges from 72\% to 94\% \citep{Kim}.  17 of the 30 subsamples had 100\% prediction success rate.  At the very worst, 3 of 11 observations were misclassified.
\section{Discussion}
In this paper, we developed two methods for Bayesian variable selection when the data has a binary outcome. The main advantage of our method compared to previous Bayesian variable selection for high-dimensional data is a vast improvement in computation time for datasets with thousands of covariates and far fewer observations. We introduced both a logistic and probit model, which allows researchers flexibility when implementing EMVS with binary data. The simulation study in Section 3.4 suggested that each model produces robust results under different circumstances. The extensive size of our simulation study shows that both methods are successful for a variety of datasets.  Furthermore, the methods presented here are available for the R statistical programming language through the library ``BinaryEMVS,'' which is available on the CRAN repository.

For both of the gene expression datasets examined above, our methods found many of the same significant genes as previous studies, but with much greater speeds. As the size of gene expression datasets continue to grow, this computational speed will make the implementation of complicated modeling feasible. Because EMVS is a relatively recent development, there are many extensions and improvements of this method available in the future. One obvious extension to our methods would involve multi-category response data. \cite{Shalev-B} introduced a Proxy-SDCA algorithm for multi-category response data that could be applied here. Another interesting extension would involve developing an EMVS model with a different hierarchical prior structure. Going forward, extensions of the EMVS model will be helpful in accommodating the increasing number of covariates needed to solve problems in a variety of areas.

\section*{Acknowledgments}
The authors would like to express their sincere gratitude to Dr. Chris Wikle, Dr. Dongchu Sun, and Dr. Sounak Chakraborty for their helpful comments during the preparation of this manuscript.

\newpage
\bibliography{reference}

\begin{thebibliography}{18}
\providecommand{\natexlab}[1]{#1}
\providecommand{\url}[1]{\texttt{#1}}
\expandafter\ifx\csname urlstyle\endcsname\relax
  \providecommand{\doi}[1]{doi: #1}\else
  \providecommand{\doi}{doi: \begingroup \urlstyle{rm}\Url}\fi

\bibitem[Ai-Jun and Xin-Yuan(2010)]{AI}
Y.~Ai-Jun and S.~Xin-Yuan.
\newblock Bayesian variable selection for disease classification using gene
  expression data.
\newblock \emph{Bioinformatics}, 26\penalty0 (2):\penalty0 215--222, 2010.

\bibitem[Albert and Chib(1993)]{AlbertChib}
J.H. Albert and S.~Chib.
\newblock Bayesian analysis of binary and polychotomous response data.
\newblock \emph{JASA}, 88:\penalty0 669--679, 1993.

\bibitem[Alon et~al.(1999)Alon, Barkai, Notterman, Gish, Ybarra, Mack, and
  Levine]{Alon}
U.~Alon, N.~Barkai, D.A. Notterman, K.~Gish, S.~Ybarra, D.~Mack, and A.J.
  Levine.
\newblock Broad patterns of gene expression revealed by clustering analysis of
  tumor and normal colon tissues probed by oligonucleotide arrays.
\newblock \emph{Proc. Natl. Acad. Sci. USA}, 96:\penalty0 6745--6750, 1999.

\bibitem[Baragatti(2011)]{Baragatti}
M.~Baragatti.
\newblock Bayesian variable selection for probit mixed models applied to gene
  selection.
\newblock \emph{Bayesian Analysis}, 6\penalty0 (2):\penalty0 209--230, 2011.

\bibitem[Ben-Dor et~al.(2000)Ben-Dor, Bruhn, Friedman, Nachman, Schummer, and
  Yakhini]{BenDor}
A.~Ben-Dor, L.~Bruhn, N.~Friedman, I.~Nachman, M.~Schummer, and Z.~Yakhini.
\newblock Tissue classicication with gene expression profiles.
\newblock \emph{Journal of Computational Biology}, 7:\penalty0 559--583, 2000.

\bibitem[Dudoit et~al.(2000)Dudoit, Fridyland, and Speed.]{Dudoits}
S.~Dudoit, J.~Fridyland, and T.P. Speed.
\newblock Comparison of discrimination methods for the classification of tumors
  using gene expression data.
\newblock \emph{Technical Report}, 2000.

\bibitem[Furey et~al.(2000)Furey, Cristianini, Duffy, Bednarski, Schummer, and
  Haussler]{Furey}
T.S. Furey, N.~Cristianini, N.~Duffy, D.W. Bednarski, M.~Schummer, and
  D.~Haussler.
\newblock Support vector machine classification and validation of cancer tissue
  samples using microarray expression data.
\newblock \emph{Bioinformatics}, 16:\penalty0 906--914, 2000.

\bibitem[George and McCulloch(1997)]{GM97}
E.~I. George and R.E. McCulloch.
\newblock 1997.
\newblock \emph{Approaches for Bayesian Variable Selection}, 7\penalty0
  (339-373), 1997.

\bibitem[George and McCulloch(1993)]{George1993}
E.I. George and R.E. McCulloch.
\newblock Variable selection via gibbs sampling.
\newblock \emph{JASA}, 88:\penalty0 881--889, 1993.

\bibitem[Golub et~al.(1999)Golub, Slonim, Tamayo, Huard, Gaasenbeck, Mesirov,
  Coller, Loh, Downing, Caligiuri, Bloomfield, and Lender]{Golub}
T.R. Golub, D.~Slonim, P.~Tamayo, C.~Huard, M.~Gaasenbeck, J.~Mesirov,
  H.~Coller, M.~Loh, J.~Downing, M.~Caligiuri, C.~Bloomfield, and E.~Lender.
\newblock Molecular classification of cancer: class discovery and class
  prediction by gene expression molecular classification of cancer: class
  discovery and class prediction by gene expression.
\newblock \emph{Science}, 286\penalty0 (531-537), 1999.

\bibitem[Gui and Li(2005)]{GuiLi2005}
J.~Gui and H.~Li.
\newblock Penalized cox regression analysis in the high-dimensional and
  low-sample size settings, with applications to microarray gene expression
  data.
\newblock \emph{Bioinformatics}, 3001-3008, 2005.

\bibitem[Kim and Cho(2004)]{Kim}
K.J. Kim and S.B. Cho.
\newblock Prediction of colon cancer using an evolutionary neural network.
\newblock \emph{Neurocomputing}, 61:\penalty0 361--379, 2004.

\bibitem[Lee et~al.(2003)Lee, Sha, Dougherty, Vannucci, and Mallick]{Lee}
E.L. Lee, N.~Sha, E.~R. Dougherty, M.~Vannucci, and B.K. Mallick.
\newblock Gene selection: a bayesian variable selection approach.
\newblock \emph{Bioinformatics}, 19:\penalty0 90--97, 2003.

\bibitem[Li et~al.(2001)Li, Weinberg, Darden, and Pederson]{Li}
L.~Li, C.R. Weinberg, T.A. Darden, and L.G. Pederson.
\newblock Gene selection for sample classification based on gene expression
  data: study of sensitivity to choice of parameters of the ga/knn method.
\newblock \emph{Bioinformatics}, 17:\penalty0 1131--1142, 2001.

\bibitem[Nguyen and Rocke(2002)]{Nguyen}
D.V. Nguyen and D.M. Rocke.
\newblock Tumor classification by partial least squares using microarrary gene
  expression data.
\newblock \emph{Bioinformatics}, 18:\penalty0 39--50, 2002.

\bibitem[Rockova and George(2014)]{rockova}
V.~Rockova and E.~I. George.
\newblock Emvs: The em approach to bayesian variable selection.
\newblock \emph{Journal of the American Statistical Association}, 109:\penalty0
  828--846, 2014.

\bibitem[Shalev-Shwartz and Zhang(2013{\natexlab{a}})]{Shalev-A}
S.~Shalev-Shwartz and T.~Zhang.
\newblock Stochastic dual coordinate ascent methods for regularized loss
  minimization.
\newblock \emph{Journal of Machine Learning Research}, 14:\penalty0 567--599,
  2013{\natexlab{a}}.

\bibitem[Shalev-Shwartz and Zhang(2013{\natexlab{b}})]{Shalev-B}
S.~Shalev-Shwartz and T.~Zhang.
\newblock Accelerated proximal stochastic dual coordinate ascent for
  regularized loss minimization.
\newblock \emph{Technical Report}, 2013{\natexlab{b}}.

\end{thebibliography}
\newpage

\section{Appendix}
\subsection{SDCA Algorithm for Binomial Data}
Let $\boldsymbol{\lambda^*}=(d^*_1,\cdots,d^*_p)^{'}$. Then,
\small
{\flushleft
\hspace{2cm} {\bf Prox-SDCA Algorithm} \\
\hspace{2cm}Initialize $w_j^{(0)}=\frac{1}{\lambda_j^* n}\sum\limits_{i=1}^{n}\alpha_i^{(0)} x_{ij}$ for each $j=1,\ldots,p$\\
\hspace{2cm}Define: $\phi^{*}(b)=b \ \text{log}(b) + (1-b) \ \text{log}(1-b)$\\
\hspace{2cm}Iterate: for $t=1,2, \dots ,T$, let $T_0=T/2$\\
\hspace{2cm}Randomly pick i\\
\hspace{2cm}$p=\boldsymbol{x}_i^\prime w^{(t-1)}$ and $q=-1/(1+e^{-p}) - \alpha_i^{(t-1)}$\\
\hspace{2cm}$s=\min \left(1, \frac{\text{log}(1+e^p)+\phi^{*}(-\alpha_i^{(t-1)})+p \alpha_i^{(t-1)}+2q^2}{q^2(4+\frac{\sum_j x_{ij}^2 \lambda_j^{*-1}}{n})}\right)$\\
\hspace{2cm}$\bigtriangleup \alpha_i = sq$\\
\hspace{2cm}$\alpha_i^{(t)}=\alpha_i^{(t-1)}+\bigtriangleup \alpha_i $ and for $k \neq i$, \ $\alpha_k^{(t)}=\alpha_k^{(t-1)}$\\
\hspace{2cm}$w_j^{(t)} \leftarrow w_j^{(t-1)} +(\lambda_j^* n)^{-1} \bigtriangleup \alpha_i x_{ij}$ for each $j=1,\ldots,p$\\
\hspace{2cm}Output $\boldsymbol{\bar{w}}=\frac{1}{T-T_0}\sum\limits_{T_0+1}^{T}\boldsymbol{w}^{(t-1)}$\\
}
\normalsize
Each time through the M-step for both logit and probit, let $\boldsymbol{\beta}^{(k+1)}=\boldsymbol{\bar{w}}$.

\subsection{E-M Steps for the Probit Model}
\begin{enumerate}
    \item E-step
        \begin{enumerate}
            \item Estimate $\boldsymbol{z}^{(k)}\vert\boldsymbol{x},\boldsymbol{y},\boldsymbol{\beta}$.
            \item Estimate $p_i^*$ and $d_i^*$.
        \end{enumerate}
    \item M-step
        \begin{enumerate}
            \item Estimate $\boldsymbol{\beta}^{(k+1)}$ using the Generalized Ridge Regression (GRR) formula:
\begin{align}
\begin{split}
\label{smw}
\boldsymbol{\beta}^{(k+1)}=(\boldsymbol{X}^\prime \boldsymbol{X}+\boldsymbol{D^*})^{-1}\boldsymbol{X}^\prime\boldsymbol{z}^{(k)}
\end{split}
\end{align} or SDCA algorithm for continuous data as described in \citet{Shalev-A}.
         \item Estimate $\boldsymbol{\theta}^{(k+1)}$ using (\ref{theta}).
        \end{enumerate}
\end{enumerate}

\subsection{plots}

\begin{figure}[H]
\centering
\includegraphics[scale=0.4]{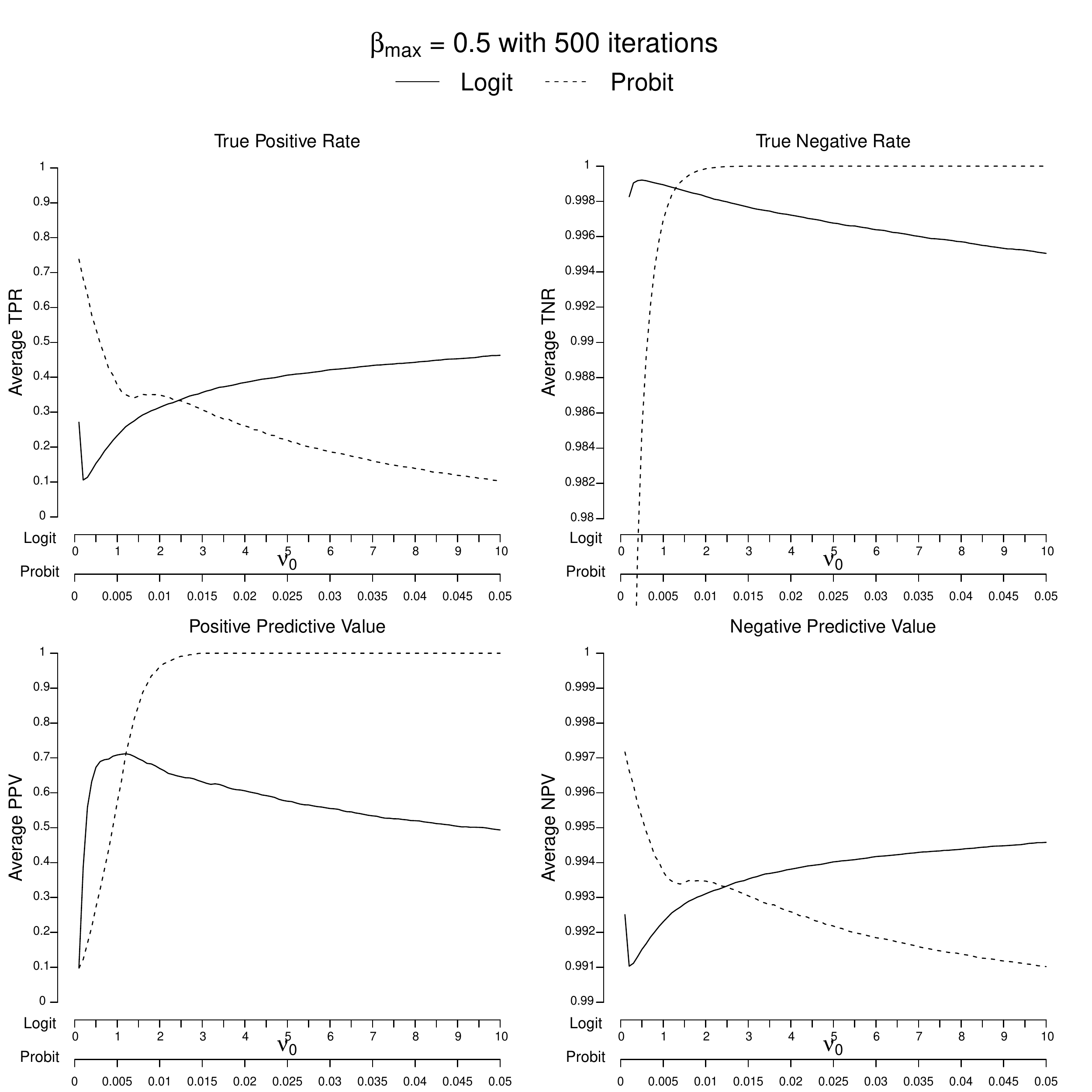}
\caption{TPR, TNR, PPV, and NPV for the logit and probit models over the tuning parameter $\nu_0$ for 500 simulated datasets using a maximum effect size of 0.5.}
\label{protbeta2}
\end{figure}

\begin{figure}[H]
\centering
\includegraphics[scale=0.4]{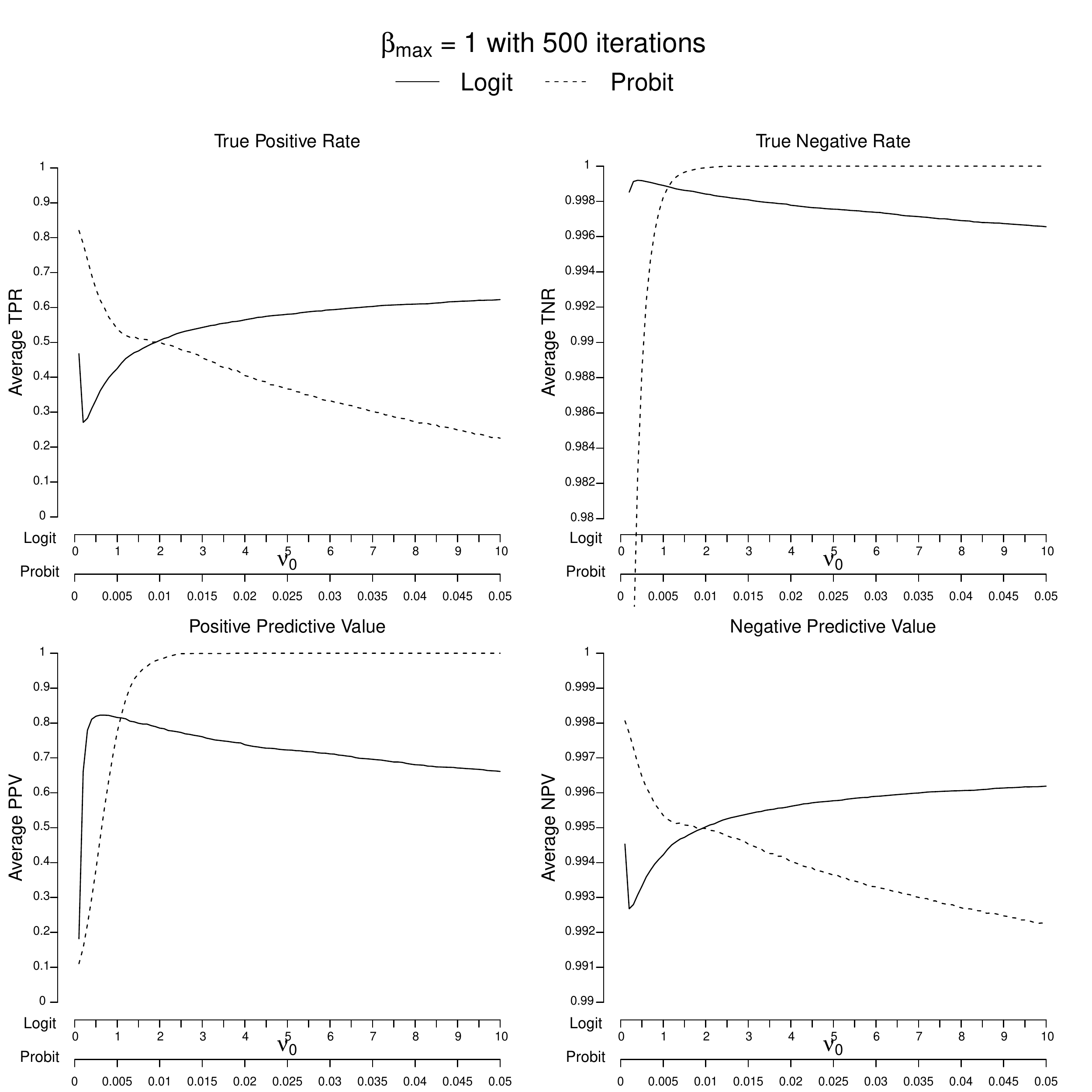}
\caption{TPR, TNR, PPV, and NPV for the logit and probit models over the tuning parameter $\nu_0$ for 500 simulated datasets using a maximum effect size of 1.}
\label{protbeta2}
\end{figure}

\begin{figure}[H]
\centering
\includegraphics[scale=0.4]{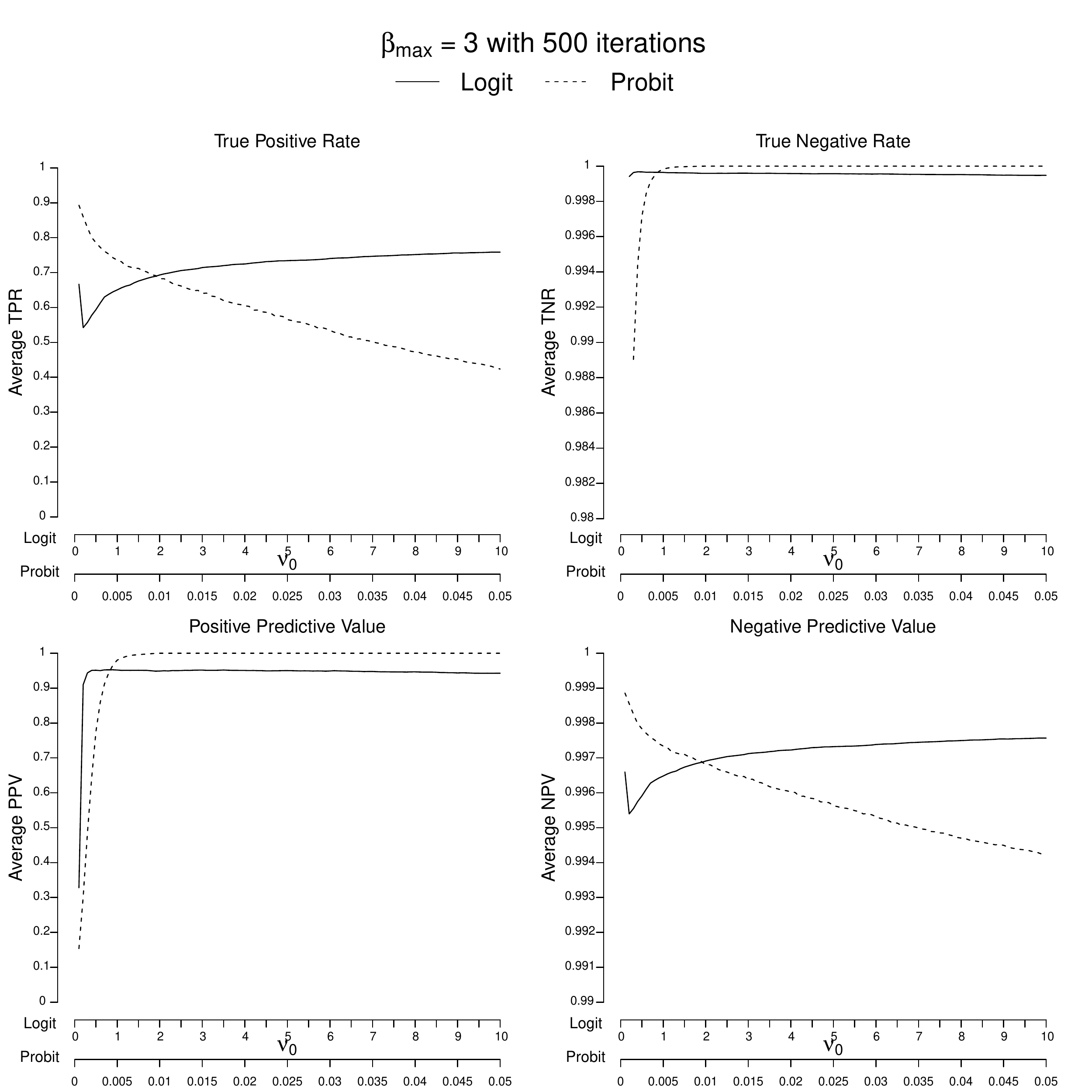}
\caption{TPR, TNR, PPV, and NPV for the logit and probit models over the tuning parameter $\nu_0$ for 500 simulated datasets using a maximum effect size of 3.}
\label{protbeta2}
\end{figure}
\end{document}